\documentclass{IEEEtran4PSCC}
\ifCLASSINFOpdf
\else
\fi
\usepackage[cmex10]{amsmath}
\usepackage{amssymb}
\usepackage{bm}

\usepackage{amsthm}
\usepackage{amsfonts}
\usepackage{booktabs}
\usepackage[colorlinks=True]{hyperref}   

\usepackage{tikz}
\graphicspath{{pics/}}

\usepackage{siunitx}
\usepackage{textcomp}

\usepackage{eurosym}

\usepackage[utf8]{inputenc}
\usepackage[english]{babel}

\usepackage{multirow}

\usepackage{footnote}
\makesavenoteenv{tabular}
\makesavenoteenv{table}

\usepackage{rotating}

\usepackage{tikz}
\usepackage{circuitikz}

\usepackage{verbatim}

\renewcommand{\baselinestretch}{0.955}
\usepackage{graphicx}

\usepackage[cmex10]{amsmath}

\usepackage{booktabs,caption,fixltx2e}
\usepackage[flushleft]{threeparttable}

\makeatletter
\let\old@ps@headings\ps@headings
\let\old@ps@IEEEtitlepagestyle\ps@IEEEtitlepagestyle
\def\psccfooter#1{%
    \def\ps@headings{%
        \old@ps@headings%
        \def\@oddfoot{\strut\hfill#1\hfill\strut}%
        \def\@evenfoot{\strut\hfill#1\hfill\strut}%
    }%
    \def\ps@IEEEtitlepagestyle{%
        \old@ps@IEEEtitlepagestyle%
        \def\@oddfoot{\strut\hfill#1\hfill\strut}%
        \def\@evenfoot{\strut\hfill#1\hfill\strut}%
    }%
    \ps@headings%
}
\makeatother

\begin{document}
	%
	% paper title
	% Titles are generally capitalized except for words such as a, an, and, as,
	% at, but, by, for, in, nor, of, on, or, the, to and up, which are usually
	% not capitalized unless they are the first or last word of the title.
	% Linebreaks \\ can be used within to get better formatting as desired.
	% Do not put math or special symbols in the title.

	%% To specify the authors when (number of affiliations <= 2)
	\author{
		\IEEEauthorblockN{Lesia Mitridati, Pascal Van Hentenryck}
		\IEEEauthorblockA{School of Industrial \& Systems Engineering \\ Georgia Institute of Technology \\
			Atlanta, Georgia \\
			\{lmitridati3,pascal.vanhentenryck\}@gatech.edu}
		\and
		\IEEEauthorblockN{Jalal Kazempour}
		\IEEEauthorblockA{Center for Electric Power and Energy  \\ Technical University of Denmark \\ Kgs. Lyngby, Denmark \\
			\{seykaz\}@elektro.dtu.dk}
			}
	
	%% To specify the authors when (number of affiliations > 2)
	% \author{\IEEEauthorblockN{Author n.1\IEEEauthorrefmark{1},
	% Author n.2\IEEEauthorrefmark{2},
	% Author n.3\IEEEauthorrefmark{3}, 
	% Author n.4\IEEEauthorrefmark{3} and
	% Author n.5\IEEEauthorrefmark{4}}
	% \IEEEauthorblockA{\IEEEauthorrefmark{1} Department Name of Organization A\\
	% Name of the organization A,
	% Address A\\ Emails if wanted}
	% \IEEEauthorblockA{\IEEEauthorrefmark{2} Department Name of Organization B\\
	% Name of the organization B,
	% Address B\\ Emails if wanted}
	% \IEEEauthorblockA{\IEEEauthorrefmark{3} Department Name of Organization C\\
	% Name of the organization C,
	% Address C\\ Emails if wanted}
	% \IEEEauthorblockA{\IEEEauthorrefmark{4}Department Name of Organization D\\
	% Name of the organization D,
	% Address D\\ Emails if wanted}
	% }
	
%	\title{An Electricity-Aware Heat Unit-Commitment Model}
	\title{A Bid-Validity Mechanism for Sequential Heat and Electricity Market Clearing}
	\maketitle

	\begin{abstract}
		Coordinating the operation of units at the interface between heat and electricity systems, such as combined heat and power plants and heat pumps, is essential to reduce inefficiencies in each system and help achieve a cost-effective and efficient operation of the overall energy system. These energy systems are currently operated by sequential markets, which interface the technical and economic aspects of the systems. In that context, this study introduces an electricity-aware heat unit commitment model, which seeks to optimize the operation of the heat system while accounting for the techno-economic interdependencies between heat and electricity markets. These interdependencies are represented by bid-validity constraints, which model the linkage between the heat and electricity outputs and costs of combined heat and power plants and heat pumps. This approach also constitutes a novel market mechanism for the coordination of heat and electricity systems, which defines heat bids conditionally on electricity prices. Additionally, a tractable reformulation of the resulting trilevel optimization problem as a mixed integer linear program is proposed. Finally, it is shown on a case study that the proposed model yields a $23\%$ reduction in total operating cost and a $6\%$ reduction in wind curtailment compared to a traditional decoupled unit commitment model.
	\end{abstract}
	

	\newtheorem{definition}{Definition}
	\newtheorem{proposition}{Proposition}
	\newtheorem{corollary}{Corollary}[proposition]
	\newtheorem{lemma}{Lemma}	
	\newtheorem{assumption}{Assumption}

\section{Introduction}

%\begin{enumerate}
%	\item The participation of combined heat and power (CHP) plants, and heat pumps at the interface between heat and electricity systems creates strong economic and physical interdependencies between them \cite{mitridati2020appendix}.
%	\item In addition: renewables create greater volatility and uncertainty in the power system: which may impact the heat system too!
%	\item Current markets should interface economic and physical aspects of the systems... But physical characteristics are not accounted for... In particular networks and unit commitment constraints
%	\item This non-coordinated operation of CHPs, heat pumps may impact their profitability, and create non cost effective operation of the systems
%\end{enumerate}

%Pressed by growing environmental concerns, 
%The rapid growth of renewable, stochastic, and non-dispatchable energy sources has increased the need for flexibility in the power system \cite{lund2007renewable,lund2015review,}.

Exploiting the synergies between the power system and other energy systems has been identified as a key component to achieve a flexible and sustainable energy future \cite{lund2007renewable,meibom2013energy,pinson2017towards}. In particular, due to its strong technical and economic interdependencies with the power system, an energy efficient and integrated heat system is central in a sustainable energy system \cite{heatroadmap,lund2010role,lund20144th}.
However, in many countries, such as Denmark, heat and electricity systems are operated by energy markets, which interface the technical and economic aspects of each system sequentially and independently. In that context, as highlighted in \cite{chen2015increasing}, the heat-driven dispatch of combined heat and power (CHP) plants and heat pumps may limit their flexibility in the electricity market and hinder the penetration of renewable energy sources. Additionally, \cite{virasjoki2018market} showed that in Nordic countries the heat-driven dispatch of CHPs in the heat market may increase their market power in the electricity system and impact electricity prices. In turn, electricity prices impact the merit order in the heat market and the profitability of CHPs.
These challenges raise an important research question: how to coordinate the operation of CHPs and heat pumps at the interface between heat and electricity systems, in order to achieve a cost-effective and efficient operation of the overall energy system in the current market environment?

From an operational point of view, this question has been tackled extensively in the literature with \textit{fully integrated} approaches. Reference \cite{chen2015increasing} proposed a combined economic dispatch for heat and electricity systems, which increases the flexibility of CHPs at the interface between both systems. In addition, \cite{li2016combined,zheng2018integrated,mitridati2018power,dai2018general} provided models of the dynamics of heat transfer in the district heating network that utilize the energy storage capacity of pipelines. These studies showed the benefits of accurately representing the techno-economic characteristics of the district heating system, in order to achieve an optimal operation of the overall energy system.
%Various studies proposed Lagrangian relaxation and Benders decomposition techniques \cite{guo1996algorithm,li2016combined,lin2017decentralized,dai2018general}, or meta-heuristic approaches \cite{mohammadi2013combined,vasebi2007combined,khorram2011harmony} to efficiently solve the combined heat and power dispatch (CHPD) problem. While these approaches provide a tractable solution method, they do not offer global optimality guarantees. On the other hand, 
%Therefore, \cite{mitridati2018power} developed a convex relaxation approach and provided a Mixed Integer Linear Programming (MILP) reformulation of the CHPD problem. An optimal solution of the relaxed problem is found, and in case of infeasibility, a method for recovering feasible solutions is devised. 
However, these approaches do not respect the current sequential and independent organization of heat and electricity markets.

By opposition with these fully integrated operational approaches, \cite{mitridati2016optimal} introduced a novel formulation of the heat market-clearing problem. This approach provided a
\textit{soft coordination} between heat and electricity systems while respecting the sequential clearing of their respective markets. This coordination is achieved by modeling the heat market clearing as a bilevel optimization problem, which seeks to minimize the heat dispatch cost in the upper level, while anticipating the impact of heat dispatch on the electricity market clearing in the lower level. Similarly, \cite{byeon2019unit} introduced the concept of \textit{bid validity} for coordinating the participation of gas-fired power plants (GFPP) in the sequential electricity and natural gas markets. This novel market mechanism enforces the validity of the bids of GFPPs in a gas-aware unit commitment problem, modeled as a trilevel optimization problem.
However, the main limitation of these approaches is their computational tractability, due to the non-convex complementarity conditions on the lower-level problems, which limits their applicability in large-scale energy systems. Additionally, although energy markets should interface the technical and economic aspects of each energy system, the economic dispatch in \cite{mitridati2016optimal} does not account for the the non-convex temperature dynamics in district heating networks, and the complex techno-economic characteristics of heat and electricity producers, such as start-up and no-load costs, and minimum on/off times. The decoupling of these operational and economic aspects may result in an inefficient or even infeasible dispatch. In order to address these challenges, \cite{byeon2019unit} proposed a convex relaxation of the complex dynamics in natural gas networks, and a tractable reformulation of the proposed trilevel optimization problem using a lexicographic function and strong duality of the middle- and lower-level problems.

In view of the state-of-the-art presented above, a gap remains in the literature to interface and coordinate the technical and economic aspects of both heat and electricity systems in the current market environment. Non-convex heat transfer dynamics can be accounted for in energy markets using a convex relaxation approach, as proposed in \cite{mitridati2018power}. However, unit commitment (UC) decisions introduce pricing challenges in energy markets and result uplift payments \cite{ruiz2012pricing,o2005efficient}. The commitment and dispatch decisions are traditionally fixed separately in European energy markets. In that context, in order to accurately represent the techno-economic interdependencies between heat and electricity systems and improve the coordination between them, the heat UC problem should anticipate the impact of the commitment of CHPs and heat pumps on both heat and electricity market-clearing problems. As a result, the contributions of this paper are threefold. 
%s
Firstly, an \textit{electricity-aware} heat UC model, formulated as a trilevel optimization problem, is proposed. In the upper level, the heat UC problem seeks to minimize the heat system operating cost, while anticipating the heat and electricity market clearings in the middle- and lower-level problems through bid-validity constraints. By opposition with fully integrated coordination approaches introduced in the literature, the proposed approach induces a soft-coordination between heat and electricity systems while respecting the sequential order of their UC and market clearing.
% Additionally, this heat UC model provides a tractable formulation of the techno-economnic characteristics of the district heating system, using a convex relaxation approach.
%
Secondly, this trilevel optimization problem is reformulated as a tractable single-level mixed integer linear program (MILP) using a lexicographic function and strong duality of the middle- and lower-level problems.
Finally, the proposed heat UC model is compared to a traditional sequential and decoupled UC model, using an IEEE 24-bus power system and 6-node district heating system. This analysis provides the basis to quantify the value of improving the coordination between heat and electricity systems.

The remainder of this paper is organized as follows. Section \ref{section:2} describes the techno-economic interdependencies between heat and electricity systems, Section \ref{section:3} introduces the proposed electricity-aware heat UC model, Section \ref{section:4} presents numerical results, and Section \ref{section:5} concludes and discusses potential extensions of this work. The extended formulations are provided in the online appendix \cite{mitridati2020appendix}.

\section{Techno-economic interdependencies between heat and electricity systems} \label{section:2}

%Heat and electricity systems have traditionally been operated independently and sequentially. Despite the decoupled operation of heat and electricity systems, the large penetration of CHPs and heat pumps at the interface between the systems yields strong economic and physical interactions. In this section, we present these existing interactions and the challenges they raise for the optimal operation of heat and electricity systems. This shows the importance of accurately modeling the techno-economic characteristics of both heat and electricity systems in order to optimally coordinate their dispatch.

\subsection{Techno-economic characteristics of CHPs and heat pumps} 

The strong linkage between the heat and electricity outputs of CHPs and heat pumps are at the interface between heat and electricity systems creates implicit interdependencies in the operation of both systems. 
Indeed, the heat and electricity outputs, $Q_{jt}$ and $P_{jt}$, of CHPs $ j \in \mathcal{I}^{\text{HP}}$ at a given time period $t \in \mathcal{T}$ are constrained by their joint Feasible Operating Region (FOR) \cite{lahdelma2003efficient}. The majority of CHPs are extraction units that can produce heat and electricity at different ratios. Their FOR can be modeled by a set of linear equations, which links their heat and electricity outputs as follows:
	\begin{subequations} \label{FOR2}
		\begin{align}
		& P_{jt} \geq  r_{j}Q_{jt}, \ \forall j \in \mathcal{I}^{\text{CHP}} , t \in \mathcal{T} \label{FOR21} \\
		& u^0_{jt} \underline{F}_{j} \leq \rho_{j}^{\text{H}}Q_{jt}+\rho_{j}^{\text{E}}P_{jt} \leq u^0_{jt} \overline{F}_{j}, \ \forall j \in \mathcal{I}^{\text{CHP}} , t \in \mathcal{T} \label{FOR22}
%		& 0 \leq Q_{jt} \leq \overline{Q}_{j} u^0_{jt} , \ \forall j \in \mathcal{I}^{\text{CHP}},  t \in \mathcal{T},  \label{FOR23}
		\end{align}
	\end{subequations}
	where  \eqref{FOR21} introduces the minimum heat to power ratio $r_j$, and \eqref{FOR22} the upper and lower bounds $\underline{F}_{j}$ and $\overline{F}_{j}$, on the fuel consumption $F_{jt} = \rho^{\text{E}}_j P_{jt} + \rho^{\text{H}}_j Q_{jt}$. Besides, $\rho^{\text{E}}_j$ and $\rho^{\text{H}}_j$ represent the electricity and heat efficiency, respectively, and $u^0_{jt} \in \{0,1\}$ the on/off state of the CHP. 
	%Eq. \eqref{FOR22} represents the maximum heat production $\overline{Q}_{j}$, such that $u^0_{jt} \in \{0,1\}$ represents the on/off state of  the CHP at each time step, with $u^0_{jt} =1$ when it is on and $u^0_{jt} =0$  when it is off. Equation \eqref{FOR21} models the minimum heat to power ratio, \eqref{FOR22} enforces the maximum fuel intake.
	%, and \eqref{FOR23} restricts the maximum heat output. 
%\begin{figure}[t]
%	\centering
%	\includegraphics[width=.4\linewidth]{FOR.png}
%	\caption{Feasible operating region of an extraction CHP}
%	\label{fig:consumer_centric_mechanism}
%\end{figure}
Additionally, the heat output $Q_{jt}$ of heat pumps $ j \in \mathcal{I}^{\text{HP}}$ is proportional to their electricity consumption $L_{jt}^{\text{HP}}$, with a fixed heat-to-power ratio represented by their coefficient of performance $\text{COP}_{j}$, such that
\begin{subequations} \label{FOR3}
	\begin{align}
	& Q_{jt} = \text{COP}_{j} L_{jt}^{\text{HP}}, \ \forall j \in \mathcal{I}^{\text{HP}} , t \in \mathcal{T}.
	%& 0 \leq Q_{jt} \leq \overline{Q}_{j}u^0_{jt},  \ \forall j \in \mathcal{I}^{\text{HP}} , t \in \mathcal{T}.
	\end{align}
\end{subequations}
%As a result, the heat market clearing model also computes bounds on the electricity production that CHPs can bid in the electricity market, as well as the electricity cosnumption of heat pumps. 

%	& \quad && \underline{P}_{jt} = \sum_{b \in \mathcal{B}^H_{jt}} \left( \overline{s}^\text{E}_{jb0t}\left( u_{jbt} - u_{j(b+1)t} \right) - \overline{P}_{jb0t} \right)  \ \forall  j \in \mathcal{I}^{\text{CHP}},t \in \mathcal{T} \label{ml13}  \\
%	& \quad && L^{\text{HP}}_{jt} = \dfrac{Q_{jt} }{ \textit{COP}_j }   \ \forall  j \in \mathcal{I}^{\text{HP}},t \in \mathcal{T} \label{ml14} 

In addition to the aforementioned physical interdependencies, the production costs of CHPs and heat pumps are intrinsically linked to their heat and electricity outputs. Indeed, as heat pumps produce heat from electricity purchased in the electricity market, their variable heat production cost $\Gamma_{jt}$ is proportional to their electricity consumption, such that
\begin{equation} \label{eq:cost0}
\begin{split}
& \Gamma_{jt}^{\text{H}} = \lambda^{\text{E}}_{nt} L_{jt}^{\text{HP}}  ,\  \ \forall n \in \mathcal{I}^\text{N} , j \in \mathcal{I}_n^{\text{HP}}, t \in \mathcal{T},
\end{split}
\end{equation}
where $\lambda^{\text{E}}_{nt}$ represents the electricity price at  node $n \in \mathcal{I}^{N}$ of the network where the heat pump $j \in \mathcal{I}_n^{\text{HP}}$ is located.
Similarly, as CHPs simultaneously produce heat and electricity from fossil fuels, their variable production cost $\Gamma_{jt}$ can be represented by a linear function of their fuel consumption, such that
	\begin{equation} \label{eq:cost1}
	\Gamma_{jt} = C_j\left(\rho^{\text{E}}_j P_{jt} + \rho^{\text{H}}_j Q_{jt}\right), \ \forall j \in \mathcal{I}^{\text{CHP}}, t \in \mathcal{T}.
	\end{equation}	
Due to the strong linkage between the heat and electricity outputs of CHPs, the cost allocation between heat and electricity production is not straightforward. A commonly used approach defines the heat production cost of CHPs $j \in \mathcal{I}_n^{\text{CHP}}$ located at node $n$ as their total production cost minus revenues from electricity sales \cite{pinson2017towards,mitridati2016optimal}, such that
\begin{equation} \label{eq:cost2}
\begin{split}
& \Gamma_{jt}^{\text{H}} = C_j\left(\rho^{\text{E}}_j P_{jt} + \rho^{\text{H}}_j Q_{jt}\right) - \lambda^{\text{E}}_{nt} P_{jt} \\
&  \ , \ \forall n \in \mathcal{I}^\text{N} , j \in \mathcal{I}_n^{\text{CHP}}, t \in \mathcal{T}.
\end{split}
\end{equation}

\subsection{Interdependencies in the current market environment} \label{section:markets}

In Nordic countries, such as Denmark, heat and electricity systems are operated by competitive auction-based markets, such as day-ahead markets, which interface the physical and economic aspects of each system. These energy markets operate on the principles of \textit{energy exchanges}, in which market participants submit bids, which implicitly embed their techno-economic characteristics. These bids are dispatched by the market operator based on a merit-order and least-cost principle. Additionally, market participants may choose to offer a certain quantity at any price, which will be dispatched regardless of the equilibrium price in the market. Contrary to energy markets operating on the principles of \textit{energy pools}, such as North American electricity markets, European markets do not explicitly account for the complex techno-economic characteristics of the market participants and the networks \cite{meeus2005development}. Furthermore, despite the aforementioned techno-economic interdependencies between heat and electricity systems, their respective markets are cleared sequentially and independently.

The day-ahead heat market is traditionally cleared prior to the day-ahead electricity market. Given its commitment status, each heat market participant $j \in \mathcal{I}^\text{H}$ submits bids $b \in \mathcal{B}^\text{H}$ for each hour $t \in \mathcal{T}$ of the following day, in the form of independent price-quantity pairs $(\alpha^\text{H}_{jbt},\overline{s}^\text{H}_{jbt})$ \cite{Varmelast}. 
The independent heat market operator clears the day-ahead heat market by minimizing the cost of dispatching these bids subject to transmission network and nodal balance constraints. The dynamics and varying time delays of heat transfer in pipelines can be modeled using a convex relaxation approach, as proposed in \cite{mitridati2018power}. However, in practice, time delays in pipelines and UC variables are fixed and the heat market operator solves an optimal heat flow problem.

Furthermore, these bids must reflect the marginal heat production cost $\dot{\Gamma}^\text{H}_{jt}$ of CHPs and heat pumps. Due to the linkage between their heat and electricity outputs and costs, CHPs and heat pumps must anticipate their day-ahead electricity dispatch $P_{jt}$ and electricity prices $\lambda^{\text{E}}_{nt}$ in order to accurately compute their marginal heat production cost. As the heat-to-power ratio of heat pumps is considered fixed and given by their COP, the marginal heat production cost of heat pumps is proportional to the foreseen electricity prices.
%, such that
%\begin{equation} \label{eq:marginal_cost1}
%\begin{split}
%& \dot{\Gamma}_{jt}^{\text{H}} = \dfrac{\lambda^{\text{E}}_{nt} }{\text{COP}_j}  \ , \ \forall n \in \mathcal{I}^\text{N} , j \in \mathcal{I}_n^{\text{HP}}, t \in \mathcal{T},
%\end{split}
%\end{equation}	
Additionally, as the heat-to-power ratio of CHPs is variable, their marginal heat production cost must be computed at the optimal heat-to-power ratio given the heat production and expected electricity prices. Therefore, for low electricity prices the marginal heat production cost of CHPs represents the incremental heat production cost at the minimum heat-to-power ratio, and for high electricity prices it represents the opportunity loss of producing an extra unit of heat at the maximum heat-to-power ratio. The formulation of these marginal heat production costs is provided in the online appendix \cite{mitridati2020appendix}.
This cost allocation method allows CHPs and heat pumps to implicitly model the linkage between their heat and electricity outputs and costs. However, the heat market clearing is myopic to the impact of the heat dispatch of CHPs and heat pumps on the electricity market clearing, and in turn, to the impact of electricity prices on the marginal heat production cost and profitability of CHPs and heat pumps \cite{virasjoki2018market}.

Once their heat dispatch for each hour of the following day is fixed, CHPs and heat pumps can participate in the day-ahead electricity market \cite{Nordpool2018bids}. Each electricity market participant $j \in \mathcal{I}^\text{E}$ places independent price-quantity bids $\tilde{b} \in \mathcal{B}^\text{E}$ for each hour of the following day, in the form $(\alpha^\text{E}_{j\tilde{b}t},\overline{s}^\text{E}_{j\tilde{b}t})$. The electricity market operator clears the day-ahead electricity market by minimizing the cost of dispatching these bids subject to transmission network and nodal balance constraints. Assuming a standard linearized DC power flow representation of the network, this market clearing can be modeled as a linear program (LP).

However, due to the strong linkage between their heat and electricity outputs and costs, the bids of CHPs and heat pumps in the electricity market are determined by their day-ahead heat dispatch $Q_{jt}$, and last selected bid $b$ in the heat market. Indeed, the minimum and maximum electricity production of CHPs, and the inflexible electricity consumption of heat pumps, are functions of their heat dispatch. Therefore, as CHPs and heat pumps are required to produce their heat dispatched, they must self-commit their minimum electricity production, and inflexible electricity consumption in the electricity market. In addition, the quantity bids of CHPs must be adjusted to be within their electricity production bounds, and therefore, are functions of the day-ahead heat dispatch and last selected bid, as detailed in the online appendix \cite{mitridati2020appendix}. However, this heat-driven dispatch limits the flexibility of CHPs and heat pumps in the electricity market \cite{chen2015increasing}.
%\begin{figure}[t]
%	\centering
%	\includegraphics[width=\linewidth]{heat_elec_bids.png}
%	\caption{Bids of a CHP in the day-ahead(a) heat market, and(b) electricity market for a given heat dispatch $Q_{jt}$ and last selected bid $b$.}
%	\label{fig:CHP_bids}
%\end{figure}

\section{Heat unit commitment model for improved coordination of heat and electricity systems} \label{section:3}

%This section describes the proposed electricity-aware heat UC model aiming at improving the coordination between heat and electricity systems. In order to streamline the notations, in the remainder of this paper, compact formulations of the heat UC, as well as heat and electricity market clearings are used.

\subsection{Electricity-awareness through bid-validity constraints}

\begin{figure}[t]
	\centering
	\includegraphics[width=.7\linewidth]{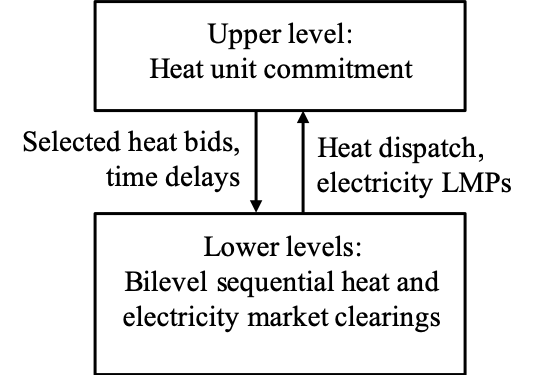}
	\caption{Trilevel electricity-aware heat unit commitment model}
	\vspace{-0.4cm}
	\label{fig:unit_commitment}
\end{figure}

As discussed in Section \ref{section:markets}, the current sequential heat and electricity market clearings do not exploit the interdependencies between the energy systems. As a result, the heat market clearing does not take advantage of the flexibility in the district heating network and is myopic to the impact of the heat dispatch on the power system. In this context, the UC decisions and time delays in the district heating network may have a strong impact on the efficient operation of both heat and electricity systems, and the profitability of CHPs and heat pumps at the interface between them.
In order to address these challenges, we propose an electricity-aware heat UC model which
%These challenges can be tackled by a heat UC problem, which optimizes the commitment decisions $u_{jbt} \in \{0,1\}$ of each heat market participant, while accounting for the interactions between heat and electricity markets, as well as the inter-temporal constraints in each system.
(i) provides an accurate representation of the techno-economic characteristics of the units at the interface between heat and electricity systems, such as CHPs and heat pumps, and the flexibility in the district heating network, and (ii) induces a \textit{soft coordination} between heat and electricity systems while respecting the order of their market clearings, as illustrated in Figure \ref{fig:unit_commitment}.

.

%%%
In practice, the techno-economic characteristics of CHPs and heat pumps, which interface heat and electricity markets, can be accounted for by bid-validity constraints. These constraints guarantee that the price of the last selected bid of each CHP and heat pump is lower than or equal to its marginal cost
\begin{equation} \label{eq:bid_validity}
\begin{split}
    & \left(\alpha^\text{H}_{jbt} - M_j\right)\left( u_{jbt} + u_{jb+1t} \right) \geq \dot{\Gamma}^\text{H}_j - M_j \\
    & \ , \ \forall j \in \mathcal{I}^\text{CHP}\cup\mathcal{I}^\text{HP}, t \in \mathcal{T}, b \in \mathcal{B}^\text{H},
\end{split}
\end{equation}
where $M_j$ is an upper bound on the marginal heat production costs of CHPs and heat pumps, and $u_{jbt} \in \{0,1\}$ are the commitment decisions on the bids submitted by CHPs and heat pumps, such that $u_{jbt}=1$ if and only if bid $b$ is selected. As the marginal heat production costs of CHPs and heat pumps can be expressed as linear functions of electricity LMPs, \eqref{eq:bid_validity} represents linear a set of constraints, which link commitment decisions and electricity LMPs. 
These constraints can also be interpreted as a market mechanism for the coordination of heat and electricity systems, which allows CHPs and heat pumps to offer heat bids which are conditional on the electricity LMPs. Indeed, in this approach each heat bid is associated with a condition on the electricity LMPs for which it is considered valid. Therefore, this bid-validity approach allows greater flexibility for CHPs and heat pumps to model their techno-economic characteristics and the linkage between their heat and electricity outputs and costs in the current sequential market design. 

Furthermore, through these bid validity constraints, the heat UC problem anticipates the impact of the commitment of CHPs and heat pumps on both heat and electricity market clearings. Therefore, this problem represents a Stackelberg game, in which the leader, i.e. the heat UC, tries to optimize its decisions, while anticipating the reaction of the followers, i.e., the sequential heat and electricity market clearings. In a Stackelberg game, the actions of the leader constrain the reaction of the follower, and, in return, the reaction of the follower impacts the objective of the leader. In this case, the commitment decisions of heat producers impact heat and electricity market clearings. In return, the heat dispatch of CHPs and heat pumps impacts the heat system operating cost in the heat UC problem. Additionally, electricity LMPs impact the marginal heat production costs and profitability of CHPs and heat pumps. This is accounted for in the heat UC problem in the form of bid-validity constraints. Assuming perfect information on the input data in heat and electricity market clearings, this problem can be modeled as a hierarchical optimization problem, where the upper-level problem, which represents the heat UC problem, is constrained by the sequential heat and electricity market-clearing problems. Due to the sequential order of their clearing and the techno-economic interdependencies between them, the heat and electricity market clearings are themselves modeled as a bilevel optimization problem. As a result, the proposed heat unit commitment model is formulated a trilevel optimization problem.

\vspace{-0.4cm}

\subsection{Upper-level problem: Heat unit commitment}

In the upper-level problem \eqref{3level}, the proposed electricity-aware heat UC problem seeks to minimize the heat system operating cost, while anticipating the impact of its decisions $\bm{z}$ on the sequential heat and electricity market clearings. Additionally, this problem receives feedback on the outcomes of the heat and electricity market clearings, i.e., heat dispatch $\bm{x^\text{H}}$ and electricity LMPs $\bm{y^\text{E}}$, in order to ensure the validity of all bids. The set of decision variables $\bm{z}$ of the heat UC problem includes the commitment, start-up and shut-down variables, and start-up costs of each heat producer, and time delays in the pipelines of the district heating network. 
%This model can formulated as follows:
%The commitment $u^0_{jt}$ of CHPs and heat pumps impact their dispatch $Q_{jbt}$ in the day-ahead heat market.
%The dispatch of CHPs and heat pumps impact the electricity production of CHPs and electricity consumption of heat pumps in the electricity market, and therefore the electricity locational marginal prices (LMPs) $\lambda^\text{E}_{nt}$.
Therefore, the electricity-aware heat UC problem can be expressed as
\begin{subequations} \label{3level}
	\begin{alignat}{7}
	& \min_{\overset{\bm{z}\in\{0,1\}^N}{\underset{\bm{x^\textbf{H}} ,  \bm{y^\textbf{E}} \geq \bm{0}}{}}} && \quad c^{0^\top}  \bm{z}  + c^{\text{H}^\top}  \bm{x^\textbf{H}} \label{3level0} \\
	& \quad \text{s.t.} && \bm{z} \in \mathcal{Z}^{\text{UC}} \label{3level1} \\
	& \quad                && \quad A^{\text{UC}}  \bm{z}                                     +     B^{\text{UC}}  \bm{y^\textbf{E}}    \geq   \ b^{\text{UC}} \label{3level2} \\
	& \quad                &&  \quad  \bm{x^\textbf{H}} , \bm{y^\textbf{E}}              \in      \text{primal and dual sol. of } \eqref{bilevel_heat_elec},  \label{3level3}
	\end{alignat}
\end{subequations}
where \eqref{3level0} represents the heat system operating cost, which includes no-load costs $\alpha_{jt}^0$, start-up costs $r_{jt}$, and the cost of dispatching the submitted bids $\alpha^\text{H}_{jbt}$, and \eqref{3level1} represents the feasible set $\mathcal{Z}^{\text{UC}}$ of UC and time-delay variables. Furthermore, the heat UC problem enforces bid-validity constraints \eqref{3level2}, and receives feedback from the middle- and lower-level problems \eqref{3level3}, representing the sequential heat and electricity markets.
%where \eqref{3level0} represents the heat system operating cost, \eqref{3level1} the unit-commiment and time delays constraints, \eqref{3level2} the bid validity copnstraints, and \eqref{3level3} the lower-level problems. 
%The matrices and vectors of parameters $c^{0}$, $A^{\text{UC}}$, $B^{\text{UC}} $,  and $b^{\text{UC}}$ can be derived from the heat UC problem presented in the online appendix \cite{mitridati2020appendix}.

\subsection{Middle- and lower-level problems: Sequential heat and electricity market clearings}

\begin{figure}[t]
	\centering
	\includegraphics[width=\linewidth]{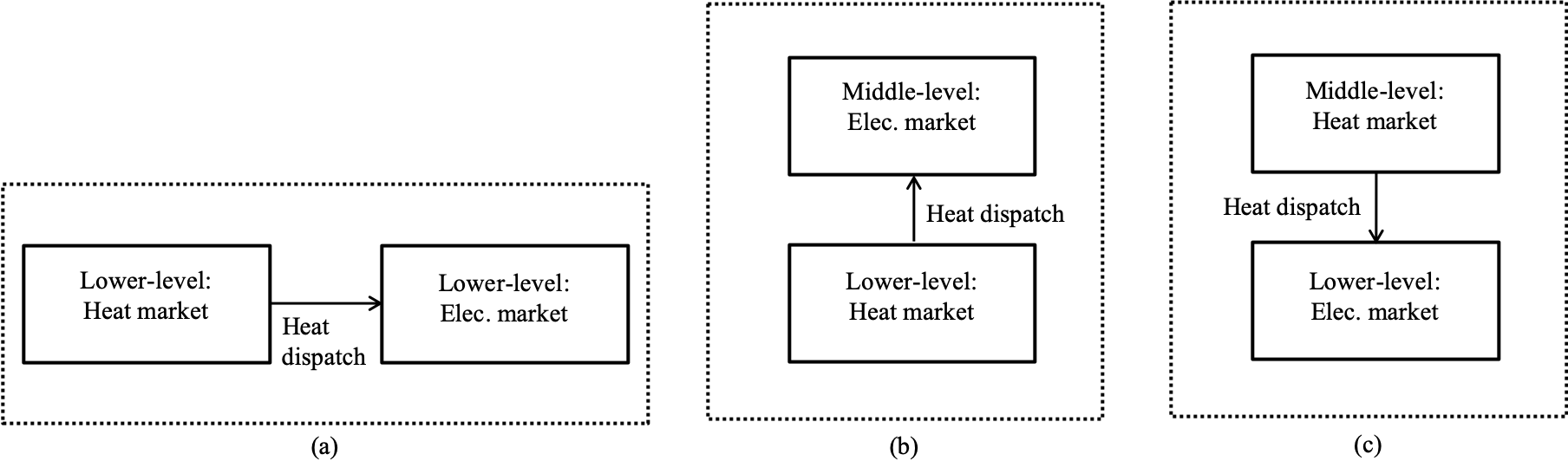}
	\caption{Formulations of the sequential heat and electricity market clearings -- (a) single-level equilibrium, (b) bilevel problem in \cite{pineda2016capacity}, and (c) proposed bilevel problem}
	\vspace{-0.4cm}
	\label{fig:3models}
\end{figure}

Despite their sequential and independent clearing, heat and electricity markets are linked through the participation of CHPs and heat pumps in both markets. Therefore, when the electricity market is cleared it is constrained by the heat dispatched in the heat market. 
As discussed in \cite{pineda2016capacity}, these sequential markets can be modeled as a single-level equilibrium problem, in which the electricity market clearing takes as input the optimal solutions of the heat market clearing, as illustrated in Figure \ref{fig:3models}(a). This approach would result in a bilevel formulation of the heat UC problem with two lower-level problems. However, representing the sequential heat and electricity market clearings as an equilibrium problem introduces non-convex bilinear terms in their dual formulation. Alternatively, \cite{pineda2016capacity} argues that these sequential market clearings can be formulated as a bilevel optimization problem, in which the leader represents the electricity market-clearing problem, which is constrained by the heat market-clearing problem, as illustrated in Figure \ref{fig:3models}(b). Additionally, this bilevel formulation can be reformulated as an LP, and would allow us to reformulate the electricity-aware heat UC as an MILP. However, if the heat market-clearing problem has multiple optimal solutions, the solution that minimizes the electricity market clearing is selected. As the heat market is myopic to the electricity market clearing, this model is not realistic. Therefore, we propose a third formulation of the sequential heat and electricity market clearings as a bilevel optimization problem, in which the middle-level problem represents the heat market clearing, which is constrained by the electricity market clearing in the lower level, as illustrated in Figure \ref{fig:3models}(c). In this model, the adjusted electricity bids are computed in the heat market clearing and considered as fixed inputs by the electricity market clearing in the lower level. 
Note that, under the assumption of the unicity of the solutions of the heat and electricity market clearings, these three models are equivalent. In case of multiple solutions of the electricity market clearing, the proposed model selects the solution that minimizes the heat UC and dispatch cost.

The set of day-ahead heat market variables $\bm{x^\textbf{H}}$ includes the dispatch of all bids, mass flow rates, temperatures, and pressures in the district heating network, as well as the adjusted electricity bids of CHPs and heat pumps. The set of variables $\bm{x^\textbf{E}}$ of the electricity market-clearing problem includes the dispatch of all bids, and the voltage angles at each node of the network. Furthermore, the vector $\bm{y^\textbf{E}}$ represents the variables of the dual electricity market-clearing problem. In particular, the dual variables associated with the nodal balance equations represent the electricity LMPs. This bilevel formulation of the sequential heat and electricity market clearings can be expressed in a compact form as follows:
\begin{subequations} \label{bilevel_heat_elec}
\begin{alignat}{7}
& \min_{\bm{x^\textbf{H}} ,  \bm{y^\textbf{E}} \geq \bm{0} }       && \quad c^{\text{H}^\top}   \bm{x^\textbf{H}}  \label{bilevel_heat_elec0} \\
& \quad    \text{s.t.}                                                                && \quad A^{\text{H}}            \left(\bm{z} \right)  	 \bm{x^\textbf{H}} + B^{\text{H}} &&    \bm{z} \geq b^{\text{H}}  \label{bilevel_heat_elec1} \\
& \quad  																				&&  \quad  \bm{y^\textbf{E}}   \in   \text{dual sol. }                                                    &&   \min_{\bm{x^\textbf{E}} \geq \bm{0}} \quad  c^{\text{E}^\top} \bm{x^\textbf{E}} \label{bilevel_heat_elec2}  \\
& \quad 																				&&  \quad                                                                                                                  &&  \  \text{s.t.} \quad A^{\text{E}}  \bm{x^\textbf{E}}   +  B^{\text{E}}  \bm{x^\textbf{H}} \geq b^{\text{E}}, \label{bilevel_heat_elec3}
\end{alignat}
\end{subequations}
%\begin{figure}[t]
%	\centering
%	\includegraphics[width=.5\linewidth]{3level.png}
%	\caption{Proposed trilevel model of electricity-aware heat UC problem}
%	\label{fig:3levels}
%\end{figure}
where \eqref{bilevel_heat_elec1} represents linearized district heating network constraints and feasibility of the heat bids. Additionally,  \eqref{bilevel_heat_elec2}-\eqref{bilevel_heat_elec3} represents the electricity market clearing, which seeks to minimize the electricity dispatch cost \eqref{bilevel_heat_elec2}, constrained by linearized DC power flow constraints and feasibility of the electricity bids \eqref{bilevel_heat_elec3}. Constraints \eqref{bilevel_heat_elec1} and \eqref{bilevel_heat_elec3} embed the interdependencies between the heat UC, heat dispatch and adjusted electricity bids of CHPs and heat pumps\footnote{The notation $A^{\text{H}}  \left(\bm{z} \right)$ represents a matrix which coefficients $a^{\text{H}}_{kl}\left(\bm{z} \right)$ are linear expressions of the upper-level variables $\bm{z}$.}.
As a result, the proposed electricity-aware heat UC formulated in \eqref{3level} is a trilevel optimization problem.

\subsection{Reformulation as a single-level optimization problem}

The proposed electricity-aware heat UC problem \eqref{3level} can be reformulated as a single-level MILP. The first step is to reformulate the middle- and lower-level problems as an LP, by noticing that the objective function \eqref{bilevel_heat_elec0} and constraints \eqref{bilevel_heat_elec1} of the middle-level problem do not depend on the lower-level primal and dual variables, $\bm{x^\text{E}}$ and $\bm{y^\text{E}}$. Therefore, the solutions of the middle-level problem are not affected by the solutions of the lower-level problem, and the bilevel sequential heat and electricity market-clearing problem \eqref{bilevel_heat_elec} can be solved in two steps: (i) solve the middle-level problem and obtain the optimal solutions $\bm{\hat{x}^{H}}$, (ii) solve the lower-level problem with $\bm{x^H}$ fixed to $\bm{\hat{x}^{H}}$ and obtain the optimal solutions $\bm{\hat{y}^\text{E}}$. Therefore, the middle- and lower-level problems can be approximated by a single-level LP using a lexicographic function \cite{byeon2019unit}. The electricity-aware UC problem can then be reformulated as a single-level MILP, by strong duality of this approximation of the middle- and lower-level problems.
\begin{proposition} \label{prop1}
The trilevel optimization problem \eqref{3level} can be asymptotically approximated by the following single-level problem:
\begin{subequations} \label{3_1level}
\begin{alignat}{2}
& \min_{\overset{\bm{z}\in\{0,1\}^N,\bm{x^\textbf{H}} \geq \bm{0} }{\underset{\bm{x^\textbf{E}} \geq \bm{0} , \bm{y^\textbf{H}},\bm{y}^\textbf{E}}{}}} &&  \gamma c^{0^\top}  \bm{z}  + \gamma c^{\text{H}^\top}  \bm{x^\textbf{H}} + \left( 1-\gamma \right) c^{\text{E}^\top}  \bm{x^\textbf{E}} \label{3_1level0} \\
& \quad \quad \text{s.t.} &&  \bm{z} \in \mathcal{Z}^{\text{UC}} \label{3_1level1} \\
& \quad &&  A^{\text{UC}}  \bm{z}  +  \dfrac{1}{(1-\gamma)} B^{\text{UC}} \bm{y^\textbf{E}}   \geq  b^{\text{UC}} \label{3_1level2} \\
& \quad       && A^{\text{H}}   \left(\bm{z} \right)  \bm{x^\textbf{H}} + B^{\text{H}}    \bm{z} \geq b^{\text{H}} \label{3_1level3} \\
 &  \quad    &&  A^{\text{E}}  \bm{x^\textbf{E}}   +  B^{\text{E}}  \bm{x^\textbf{H}} \geq b^{\text{E}} \label{3_1level4} \\
& \quad       && \bm{y^{\textbf{H}^\top}}  A^{\text{H}}  \left(\bm{z} \right) + \bm{y^{\textbf{E}^\top}} B^{\text{E}}   \leq \gamma c^{\text{H}^\top} \label{3_1level5} \\
 &  \quad    &&  \bm{y^{\textbf{E}^\top}} A^{\text{E}}  \leq \left( 1-\gamma \right) c^{\text{E}^\top} \label{3_1level6} \\
 & \quad &&  \bm{y^{\textbf{H}^\top}}   \left( b^{\text{H}} - B^{\text{H}}    \bm{z}  \right)  + \bm{y^{\textbf{E}^\top}} b^{\text{E}}  \nonumber \\
 & \quad && \ \geq  \gamma c^{\text{H}^\top}  \bm{x^\textbf{H}} + \left( 1-\gamma \right) c^{\text{E}^\top}  \bm{x^\textbf{E}}.  \label{3_1level7} 
\end{alignat}
\end{subequations}
When the penalty factor $\gamma$ tends to 1, the solutions of \eqref{3_1level} converge to the solutions of \eqref{3level}.
\end{proposition}
The proof of Proposition \ref{prop1} is provided in the online appendix \cite{mitridati2020appendix}. Note that the bilinear terms in \eqref{3_1level3}, \eqref{3_1level6}, and \eqref{3_1level7} can be linearized using an exact McCormick relaxation \cite{mccormick1976computability}. 

%\section{Benders decomposition algorithm}

\section{Numerical results} \label{section:4}

\subsection{Case study setup}

We compare the proposed electricity-aware heat UC model to a traditional decoupled heat UC model, which seeks to minimize the heat UC and dispatch cost \eqref{3level0}, constrained by UC and time delays constraints \eqref{3level1}, as well as linearized district heating network constraints and feasibility of the heat bids \eqref{bilevel_heat_elec1}. Contrary to the proposed electricity-aware heat UC model, this decoupled UC model is myopic to the impact of the commitment of CHPs and heat pumps on the electricity market, and in turn the impact of electricity prices on the feasibility of heat bids. For simplicity, and in order to compare both models hour-by-hour, time delays in pipelines are fixed at zero in the heat UC problem. The impact of varying time delays is discussed in details in \cite{mitridati2018power}.

These two UC models are implemented on a 24-bus power system, and two isolated 3-node district heating systems representative of the Danish energy system. The modified version of the 24-bus IEEE Reliability Test System consists of twelve thermal power plants, six wind farms, two extraction CHPs, and two heat pumps. Data for power generation, costs, transmission, and loads for the 24-bus IEEE Reliability Test System is derived from \cite{ordoudis2016updated}. Additionally, spatially and temporally correlated profiles of wind power generation at six locations are derived from \cite{Bukhsh2015}. 
Each district heating network consists of one CHP, one waste incinerator heat-only (HO) unit, one heat-only peak boiler, and one large-scale heat pump each. The techno-economic characteristics of these units and district heating transmission networks, and heat load profiles are derived from \cite{zugno2016commitment,li2016combined,mitridati2018power}, and \cite{Madsen2015}. Although these two district heating networks are similar, the main difference between them, which is expected to lead to different UC and dispatch, results from their interdependencies to the power system at different nodes.
Indeed, the heat bids and no-load costs of CHPs and heat pumps are computed using the average electricity LMPs at the nodes of the power system to which they are connected \cite{ordoudis2016updated}.
At day-time hours, i.e. between hours $7$ to $20$, the average LMPs are $\$5.9$/MWh in the first district heating network, and $\$11.3$/MWh in the second district heating network. Besides, it is $\$0$/MWh at night-time hours in both district heating networks. The technical data about heat and electricity networks, load profiles, wind generation, and price-quantity bids is provided in the online appendix \cite{mitridati2020appendix}. Finally, the penalty factor $\gamma$ in  \eqref{3_1level} is fixed to $0.99$. A sensitivity analysis reveals that solutions are stable around this value, and therefore, are assumed to have converged.

\subsection{Results}

\begin{figure}[t]
	\centering
	\includegraphics[width=\linewidth]{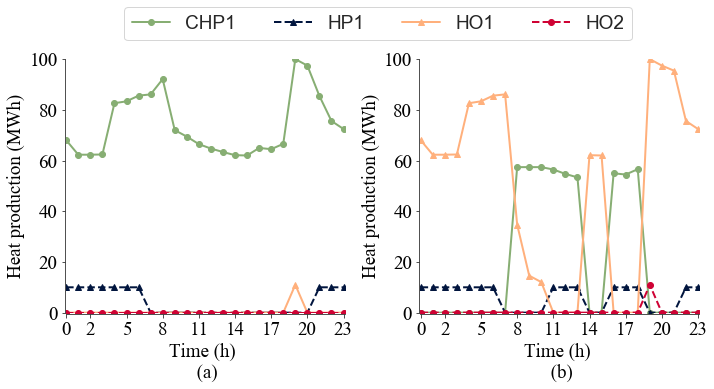}
	\caption{Heat dispatch in the first district heating network with (a) decoupled, and (b) proposed electricity-aware heat unit commitment models}
	\label{fig:Q_1}
\end{figure}

\begin{figure}[t]
	\centering
	\includegraphics[width=\linewidth]{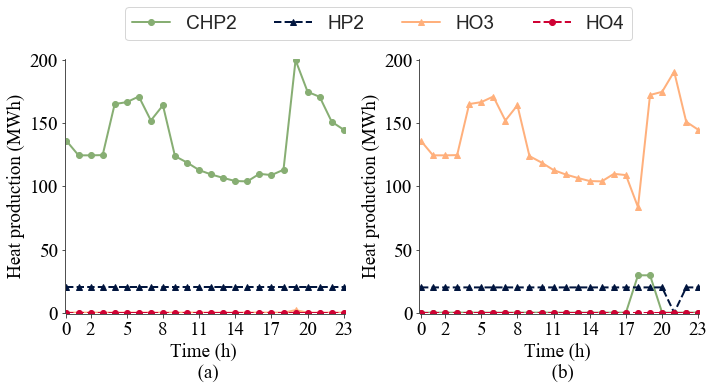}
	\caption{Heat dispatch in the second district heating network with (a) decoupled, and (b) proposed electricity-aware heat unit commitment models}
	\vspace{-0.4cm}
	\label{fig:Q_2}
\end{figure}

As expected, the heat dispatch in both district heating networks are vastly different, despite their identical techno-economic characteristics, due to their interdependencies with the power system, as illustrated in Figures \ref{fig:Q_1} and \ref{fig:Q_2}. Indeed, the difference in electricity LMPs between the nodes of the power system to which the district heating networks are connected result in different bids and merit orders. In the first district heating network, due to low average electricity prices at night-time hours, the heat pump $\text{HP}_1$ offers at the lowest prices, followed by $\text{CHP}_1$, the waste incinerator $\text{HO}_1$, and the peak heat boiler $\text{HO}_2$. Moreover, due to high average electricity LMPs at day-time hours, $\text{CHP}_1$ is cheaper than $\text{HP}_1$. Whereas, in the second district heating network, due to lower average electricity LMPs, the heat pump $\text{HP}_2$ offers at the lowest prices at each hour of the day, followed by $\text{CHP}_2$, the waste incinerator $\text{HO}_3$, and the peak heat boiler $\text{HO}_4$. These merit orders are directly reflected in the heat dispatch of the decoupled model in Figures \ref{fig:Q_1}(a) and \ref{fig:Q_2}(a). However, this myopic heat UC and dispatch results in electricity LMPs which do not support cost-recovery for $\text{CHP}_1$ at night-time hours and between hours $14$ and $15$, and for $\text{CHP}_2$ at each hour of the day. Indeed, during these hours their true heat marginal costs, computed based on the realized electricity LMPs, are higher than their submitted bids. This results in a financial loss of $\$4,389$ for $\text{CHP}_1$ and $\$5,620$ for $\text{CHP}_2$.
%\begin{figure}[t]
%	\centering
%	\includegraphics[width=\linewidth]{results_elec_prices_IEEE_24_n8_n15.png}
%	\caption{Electricity LMPs at nodes $n_8$, and $n_{15}$ with(a) DC optimal power flow on the 24-bus IEEE Reliability Test System,(b) decoupled heat UC, and(c) electricity-aware heat UC}
%	\label{fig:prices}
%\end{figure}
By anticipating the impact of the commitment and dispatch of these CHPs on the electricity market clearing and LMPs, the proposed electricity-aware model rejects the bids of $\text{CHP}_1$ and $\text{CHP}_2$ during these hours, and switches heat production to the waste incinerators $\text{HO}_1$ and $\text{HO}_3$, as observed in Figures \ref{fig:Q_1}(b) and \ref{fig:Q_2}(b). This bid-validity approach provides a better representation of the techno-economic characteristics of CHPs and heat pumps at the interface between heat and electricity systems, which ensures their profitability in the current sequential market design.

\begin{figure}[t]
	\centering
	\includegraphics[width=\linewidth]{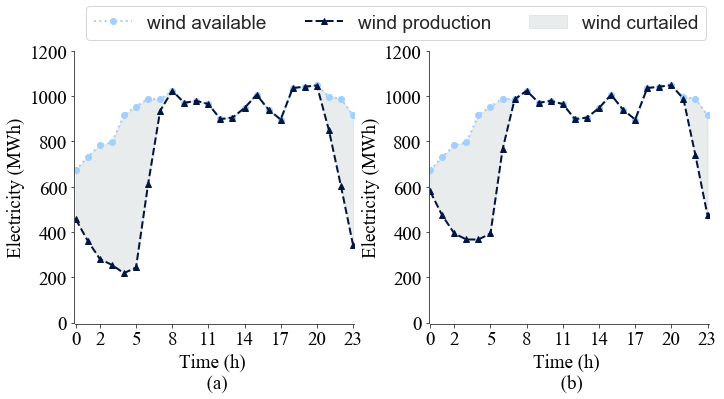}
	\caption{Wind production and curtailment with (a) decoupled, and (b) proposed electricity-aware heat unit commitment models}
	\label{fig:wind}
\end{figure}

Additionally, this electricity-aware heat dispatch provides greater operational flexibility to CHPs and heat pumps in the electricity-market, which results in higher wind utilization. In particular, the commitment and dispatch of $\text{CHP}_1$ and $\text{CHP}_2$ at night-time hours, in the decoupled model results in high wind curtailment, as illustrated in Figure \ref{fig:wind}. By switching heat production to the waste incinerators $\text{HO}_1$ and $\text{HO}_3$, the proposed model reduces wind curtailment by $6\%$ over the day.
%Indeed, as highlighted in Figure \ref{fig:prices} the resulting LMPs with the decoupled model do not support cost-recovery for $\text{CHP}_1$ during the day, whereas the proposed model anticipates on these LMPs through bid-validity constraints.

As a result, the proposed model reduces the operating cost in the overall energy system by $23\%$ and in the power system by $23\%$ compared to the decoupled model, as summarized in Table \ref{table:costs}. Besides, the electricity-aware heat UC model rejects heat bids with low prices because they violate bid-validity constraints, i.e., the submitted bids are lower than the units' foreseen marginal heat production costs. Therefore, as expected, the proposed model achieves a higher heat system operating cost, which is computed based on the submitted bids, than the decoupled model.
\begin{table}[t]
%	\resizebox{.5\textwidth}{!}{
  \begin{threeparttable}
\captionsetup{font=footnotesize}
	\caption{Heat, electricity and overall energy systems commitment and dispatch costs, in $10^3 \$$, and wind curtailment in the power system, as a $\%$ of the available wind production. }
	\label{table:costs}
	\centering
		\begin{tabular}{lccc}
			\hline
			& & \textbf{Decoupled} & \textbf{Electricity-aware} \\
			%\textbf{Heat market clearing*} & $-$ & $5,975$ & $11,129$ \\
			\textbf{Overall system cost$^*$} & $(10^3 \$)$ & $181.6$ & $139.5$ \\
			\textbf{Heat system cost}  & $(10^3 \$)$ & $85.6$ & $103.3$ \\			
			%\textbf{Electricity market clearing*} & $-$ & $52,624$ & $46,539$ \\
			\textbf{Electricity system cost} & $(10^3 \$)$ & $97.4$ & $37.9$ \\
			\textbf{Wind curtailment} & $(\%)$ & $20.4$ & $14.3$ \\
			\hline
	\end{tabular}
    \begin{tablenotes}
      \small
      \item $^*$ For consistency, the bids of heat pumps are not accounted for in the heat UC cost, as they are implicitly reflected in the electricity market-clearing cost.
    \end{tablenotes}
  \end{threeparttable}
  \vspace{-0.4cm}
\end{table}

\section{Conclusion} \label{section:5}

This paper provides a novel electricity-aware heat UC model that improves the coordination between heat and and electricity systems through bid-validity constraints, while respecting the current sequential market design. This approach can also be understood as a market-mechanism for the coordination of heat and electricity systems, which allows CHPs and heat pumps to offer heat bids which are conditional on the electricity LMPs.
Additionally, a tractable MILP reformulation of the resulting trilevel optimization problem is proposed. Finally, the value of improving the coordination between heat and electricity systems is illustrated in a case study. This simulation shows that the proposed model is able to ensure cost recovery for CHPs and heat pumps in the heat market, while reducing the operating cost of the overall energy system by $23\%$ and wind curtailment by $6\%$ compared to a decoupled UC model.
This study opens up various opportunities for future work. Firstly, the computational burden of the proposed model can be addressed using a Benders decomposition approach, as introduced in \cite{byeon2019benders}. 
%Furthermore, the proposed model provides an \textit{optimistic} hierarchical optimization problem, i.e., if multiple solutions of the lower-level problems exist, the one that minimizes the upper level objective ius selected. In order to relax this restrictive assumption, a \textit{pessimistic} trilevel optimization problem could be investigated \cite{dempe2014necessary}. However, such problems are challenging to solve and advanced optimization tools must be developed. 
Secondly, the proposed model assumes perfect information on the bids of the market participants and wind availability in the electricity market clearing. However, such information may not be communicated by the independent market operator, due to privacy concerns. Therefore, imperfect information on the parameters of the lower-level problem may be assumed using a scenario-based stochastic programming framework, or a robust counterpart of the middle- and lower-level problems. Finally, a privacy-preserving extension of the proposed model can be developped, where differential privacy on the electricity bids is ensured while preserving feasibility of the market clearing, as discussed in \cite{fioretto2018constrained}.

\section*{Acknowledgement}
This research is partly supported by an NSF CRISP Award (NSF-1638331). The authors are grateful to Prof. Pierre Pinson for discussions on early versions of this work.

\bibliographystyle{IEEEtran}	
\bibliography{mybib}

\end{document}